\documentclass[runningheads]{svmult}

\usepackage{makeidx}   
\usepackage{graphicx}  
\usepackage{subeqnar}  
\usepackage{multicol}  
\usepackage{physprbb}  
\makeindex             

\begin{document}
%
\title*{The cosmic infrared background (CIRB) and the role of the
``local environment of galaxies'' in the origin of present-day stars}
\toctitle{The cosmic infrared background and the role of the ``local
galaxy density'' in the origin of present-day stars}
%
%
\titlerunning{Cosmic infrared background}
%
\author{David Elbaz\inst{1}
\and Delphine Marcillac\inst{1}
\and Emmanuel Moy\inst{1}}
\authorrunning{Elbaz, Marcillac, Moy}
%
%
\institute{CEA Saclay$/$DSM$/$DAPNIA$/$Service d'Astrophysique,
Orme des Merisiers, F-91191 Gif-sur-Yvette Cedex, France}

\maketitle              

\begin{abstract}
A combination of evidence is presented suggesting that the majority of
the stars in today's galaxies were born during a luminous infrared
phase (LIRP) triggered by the local environment of galaxies. The CIRB
is a fossil record of these LIRPs and therefore reflects the influence
of triggered star formation through galaxy-galaxy interactions,
including non merging tidal encounters.  This scenario, in which
galaxies experienced several LIRPs in their history, is consistent
with the measured redshift evolution of the cosmic density of star
formation rates and of stellar masses of galaxies.
\end{abstract}

\section{Introduction}
Stars represent only 15\,\% of the cosmic baryonic density, itself
only equal to about 4\% of the critical density ($\Omega_{\rm
b}\simeq$ 0.04), and are unequally distributed into spheroids (10\% of
$\Omega_{\rm b}$, including spiral bulges) and disks (5\% of
$\Omega_{\rm b}$, Fukugita, Hogan \& Peebles). It is usually assumed
that disk stars formed quiescently while bulge stars formed more
efficiently and rapidly, as suggested by their redder colors and
overabundance in $\alpha$-element over iron ratio, typical of a SNII
origin. However recent studies of the history of the star formation of
the disk of the Milky Way indicate that during the last 2 Gyr it has
experienced about five major events of star formation, starbursts,
whose signatures can be found in the peaked ages of these open
clusters (de la Fuente Marcos \& de la Fuente Marcos 2004, and
references therein). As a result we may wonder whether quiescent star
formation did play a major role in the formation of present-day stars
at all. There are some evidence that star formation mostly takes place
in globular clusters and is rarely isolated. These clusters are
thought to evolve into unbound stellar associations, which evolve and
dissolve in a time-scale of about 50 Myr. This timescale is longer
than the lifetime of massive stars which dominate the luminosity of
starbursting galaxies or regions of galaxies. Hence, it is logical to
expect that if star formation occured mainly in starburst episodes
within galaxies, then the bulk of galaxies luminosity will be absorbed
by dust in the giant molecular clouds, while if star formation is
quiescent then only a minor fraction of a galaxy's luminosity will be
affected by dust extinction. We will argue in the following that there
is presently a solid collection of evidence suggesting that most stars
that we see in the local universe formed during starburst episodes.

A major peace of evidence for that comes from the detection of a
strong diffuse cosmic infrared background (CIRB, Puget et al. 1996,
Hauser \& Dwek and references therein) which is majoritarily produced
by luminous infrared phases (LIRP) within galaxies located around
$z\sim$ 0.7, for the peak of the CIRB at $\lambda \sim$ 140\,$\mu$m,
while the $\lambda \geq$ 240 \,$\mu$m is due to galaxies at $z\sim$ 2
and above (Elbaz et al. 2002, Chary \& Elbaz 2001). We have introduced
the term LIRP instead of the classical one, luminous and
ultra-luminous infrared galaxies, i.e. LIRGs and ULIRGs, because we
wish to emphasize the idea that the scenario that is emerging from the
study of distant galaxies is that LIRGs do not represent a population
of galaxies that would require to be studied independantly in order to
determine which present-day galaxies are the remnants of these LIRGs,
but what is suggested instead is that they illustrate the omnipresence
of rapid and efficient star formation as a leading process in shaping
the present-day universe, i.e. that any galaxy that we see today must
have experienced a phase when it radiated the bulk of its light in the
infrared (see also Elbaz \& Cesarsky 2003). This phase should not be
restricted to LIRGs and ULIRGs, i.e. galaxies with infrared (IR)
luminosities larger than 10$^{11}$ L$_{\odot}$ or star formation rates
(SFR) larger than $\sim$ 20 M$_{\odot}$ yr$^{-1}$, since the closest
starburst M82, for example, presents a spectral energy distribution
typical of most LIRGs, with the bulk of its luminosity radiated in the
IR although its luminosity is only 4$\times$10$^{10}$ L$_{\odot}$.

We present evidence suggesting that the bulk of local stars formed
during a LIRP. A spectroscopic diagnostic is used to quantify the
typical duration of this phase and the amount of stars that formed
during it. From the combination of both we will advocate that not only
did all galaxies experience a LIRP but that they must have experienced
several of them. Finally we will discuss the physical origin of the
LIRP and present evidence that a major event in the lifetime of
galaxies was probably underestimated: the effect of the ``local
environment of galaxies'' (LEG) and its impact in terms of driving the
conversion of gas into stars through passing-by galaxies.

\section{Luminous IR phases and the origin of present-day stars}
The detection of a CIRB came as a surprise since local galaxies only
radiate $\sim$ 30\,\% of their bolometric luminosity in the mid to far
IR range, i.e. sharing as a common origin stellar photons reprocessed
by dust above $\lambda\sim$ 3\,$\mu$m. With about half of the diffuse
background light radiated above and below this wavelength cutoff, the
extragalactic background light tells us that in the past, major star
formation events were strongly affected by dust extinction even when
galaxies were less metal rich. Deep surveys in the mid infrared
($\lambda\sim$ 15\,$\mu$m with ISOCAM onboard ISO, Elbaz et al. 1999)
brought independant evidence that infrared was more ubiquitous in the
past. These surveys detected ten times more objects at faint flux
levels than expected from the extrapolation of the local universe (no
evolution models). These galaxies turned out to belong to the class of
LIRGs and ULIRGs discovered by IRAS in the local universe but located
at a median redshift of z$\sim$0.7.  They do not exhibit any optical
signature of such strong SFRs neither in their optical colors nor in
their emission lines, except if careful correction for extinction is
applyied using the Balmer decrement (Cardiel et al. 2003, Flores et
al. 2004). Only less than 20\,\% of them were found to harbor or be
dominated by an active galactic nucleus (AGN; see Fadda et
al. 2002). Unexpectedly, because of the complex and multiple physical
origins of the mid and far IR photons (PAHs, Very Small Grains, Big
Grains), local galaxies do exhibit a strong correlation between their
mid and far IR luminosities over three decades in luminosity,
including the extreme LIRGs and ULIRGs (Elbaz et al. 2002). When
applyied to galaxies up to z$\sim$1 these correlations can be used to
derive far IR luminosities, i.e. SFRs, which are consistent with those
derived from the radio, using the radio-far IR correlation, suggesting
that these correlations are still valid at these redshifts (Elbaz et
al. 2002).

\begin{figure*} 
\begin{center}
\includegraphics[width=1.\textwidth]{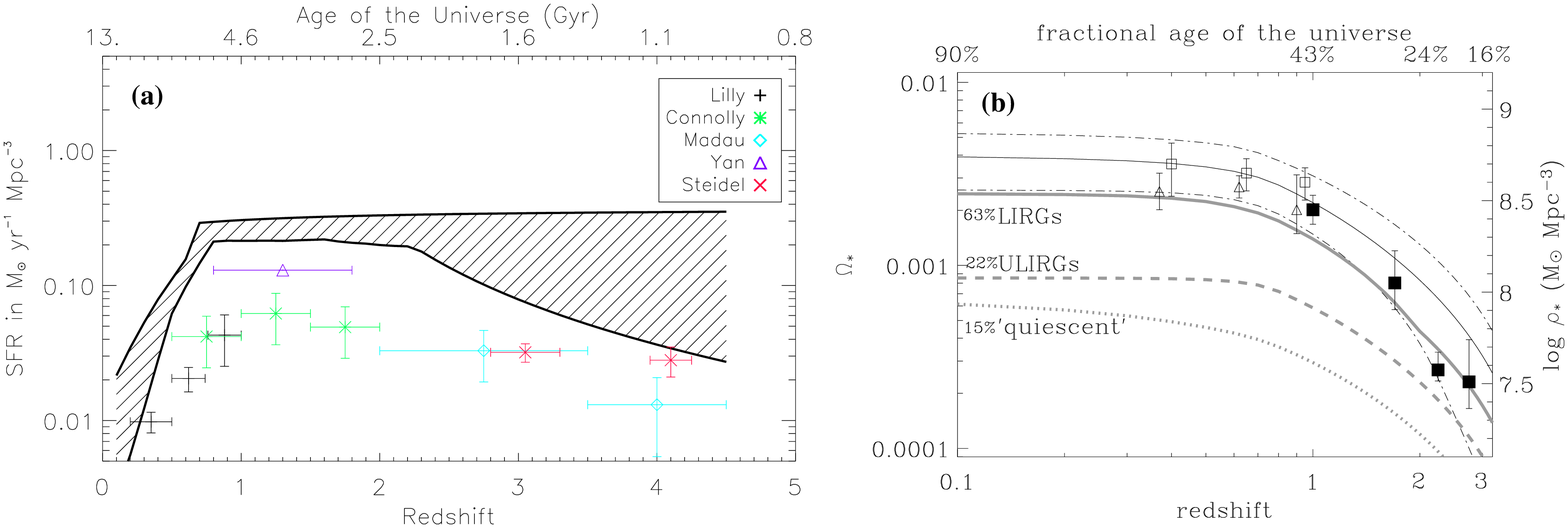}
\end{center}
\caption{{\bf a)} cosmic star formation rate (CSFR) as a function of
redshift and universe age, in a $H_0$ = 50 km s$^{-1}$ Mpc$^{-1}$ and
$q_0$ = 0.5 cosmology (Fig.15 of Chary \& Elbaz 2001). The data
represent the SFR density derived from UV or H$\alpha$ uncorrected for
extinction (the various authors are quoted on the plot and the
references can be found in Chary \& Elbaz 2001). {\bf b)} cosmic
stellar mass history (CSMH) or redshift evolution of $\Omega_{\star}$,
the cosmic stellar mass density over critical density (cosmology:
$\Lambda$=0.7, $\Omega_{\rm m}$=0.3, $H_0$ = 70 km s$^{-1}$
Mpc$^{-1}$). The data are from Dickinson et al. (2003). See text for
description.}
\label{FIG:CSFH}
\end{figure*}

The median SFR of the galaxies responsible for the evolution of the
mid IR counts is $\sim$ 50 M$_{\odot}$ yr$^{-1}$ and their
contribution to the CIRB from 100 to 1000\,$\mu$m is derived to be as
large as two thirds of its measured intensity. Hence the bulk of the
CIRB results from LIRPs at redshifts of the order of z$\sim$1, but due
to cosmological dimming the influence of more distant galaxies is more
limited to the large wavelength tail of the CIRB. Some models have
been designed to reproduce number counts in the mid IR (ISOCAM,
15\,$\mu$m), far IR (ISOPHOT-90, 175\,$\mu$m) and sub-mm (SCUBA,
850\,$\mu$m) together with the CIRB which can be used to derive the
cosmic star formation history of the universe (CSFH) unaffected by
dust extinction and to disentangle the relative roles of rapid (LIRPs)
and quiescent star formation. In Chary \& Elbaz (2001), we suggested
to define a region delimiting all possible histories of star formation
(see Fig.~\ref{FIG:CSFH}a), instead of a single line for any favorite
model that would not represent the uncertainties of the model and
observational constraints. One way to check the robustness of such
models consists in comparing to direct measurements the resulting
cosmic stellar masses history (CSMH), i.e. the redshift evolution of
$\Omega_{\star}$, the mass density of stars per comoving volume over
the critical density of the universe. The hatched area of
Fig.~\ref{FIG:CSFH}a which represents the range of possible CSFH was
converted into a range of possible CSMH in Fig.~\ref{FIG:CSFH}b, with
the thin plain line showing the mean value and the range of possible
histories delimited by the two dot-dashed thin lines. The model fits
the data collected in Dickinson et al. (2003), where stellar masses
were directly measured from optical-near IR magnitudes. 

Before deriving any conclusion, we wish to remind our assumptions: the
CSFH was derived assuming mid to far IR correlations similar to
locally (in agreement with the radio), the AGN fraction was supposed
to make a minor contribution (see above), we assumed a universal IMF
(we used the one of Gould et al. 1996, which combines a Salpeter IMF
with the now standard inflexion of the IMF below 1 M$_{\odot}$). If
these assumptions are indeed justified, then the fit of the data in
Fig.~\ref{FIG:CSFH}b illustrates the fact that the photons emitted by
star forming regions do reflect the stellar mass building of galaxies
and as a consequence, it is now possible to derive which fraction of
present-day stars were formed in a LIRP. The model CSFH was separated
into three components shown as thick grey lines in the
Fig.~\ref{FIG:CSFH}b, with 63\,\% of present-day stars born during an
IR phase of the LIRG type, which would be the dominant mode of star
formation in the universe. The shape of the redshift evolution of
$\Omega_{\star}$ implies that $\sim$80\,$\%$ of present-day stars were
born below z=2, and $\sim$50\,$\%$ below z=1, most of which during a
LIRP, leaving little room for quiescent star formation.

If most of today's stars formed during a LIRP and if this phase
reflects efficient and rapid star formation then this suggests that
some ``positive feedback'', i.e. triggering, might be at play. This
idea is comforted by the morphology and local environment (LEG) of
distant LIRGs. It is well-known that galaxies lie in large-scale
structures made of walls, filaments and clusters but LIRGs tend to
appear exclusively in high density environments around z$\sim$0.7. The
deepest ISOCAM survey was performed in a region of 27$'^2$ centered on
the Hubble Deep Field North (HDFN), detecting 95 galaxies among which
47 lie above the completeness limit of $\sim$ 0.1 mJy.  

\begin{figure*} 
\begin{center}
\includegraphics[width=1.\textwidth]{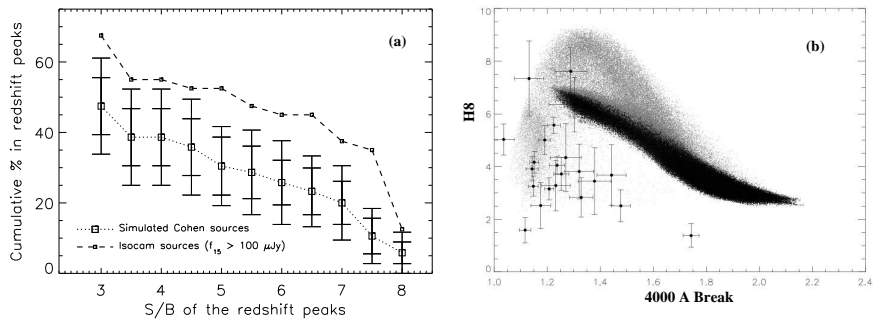}
\end{center}
\caption{{\bf a)} Cumulative fraction of galaxies located in a
redshift peak of a given S/N, i.e. number of galaxies in all redshift
peaks above a given S/N (see text). {\bf b)} Equivalent width of the
high-order Balmer absorption line, H8, as a function of the 4000\,\AA
break. Data with error bars are a sub-sample of $z\sim$ 0.7
LIRGs. Dots were generated by a Monte Carlo simulation of 200,000
galaxies. The darkest region is populated by
galaxies dominated by continuous star formation, or bursts older than
$\sim$ 2 Gyr.}
\label{FIG:zpeakH8}
\end{figure*}

The histogram of field galaxies presents several redshift peaks. The
location and the extent of these peaks can be quantified by adopting a
tresholding S/N ratio of 3, where the "background" is simply the
redshift distribution smoothed with a gaussian with $\sigma$ = 15,000
km s$^{-1}$. Monte-Carlo simulations performed by extracting 100
random samples from the real distribution of field galaxies show that
the probability of reproducing by chance the level of clustering of
the ISOCAM galaxies is much less than one chance over ten (largest
error bars). These results are illustrated on Fig.~\ref{FIG:zpeakH8}a,
where we plot the fraction of ISOCAM galaxies included in redshift
peaks above a given S/N ratio as a function of this S/N. The
correponding curves for the field galaxies and the simulated samples
are also indicated. The thin (resp. thick) error bars show the 68 \%
(resp. 90 \%) confidence level. The strong clustering of ISOCAM
galaxies illustrates that at z$\sim$0.7, galaxies experiencing a LIRP
are more clustered on average than field galaxies. On the contrary,
the study of LIRPs in the local universe using the shallower ELAIS
survey (Oliver et al., in these proceedings) shows that they are less
clustered than field galaxies locally. The natural explanation for
this behavior is that galaxies might experience a LIRP when located in
a region which is collapsing over a large scale, i.e.  star formation
would be triggered by large-scale structures in the process of
formation.

{\it How much stellar mass does a LIRG form ?} We addressed this
question using an optical spectroscopic diagnostic combining the
equivalent width of the high order Balmer absorption line H8 to the
4000\,\AA break to caracterize these starburst events (Marcillac et
al., in prep.). The advantage of using H8 instead of H$\delta$ is that
it is in a bluer side of the spectrum, hence less affected by sky
lines, and that its underlying nebular emission line can be neglected.
Fig.~\ref{FIG:zpeakH8}b shows a sub-sample of 22 LIRGs, at z$\sim$0.7,
selected in three different locations of the sky compared to a Monte
Carlo simulation of 200,000 galaxies using the code of Bruzual \&
Charlot (2004) such as those used by Kauffmann et al. (2003) to
reproduce the behavior of local galaxies in the Sloane survey. This
simulation can be used to determine which histories of star formation
would reproduce these galaxies. It is found that only galaxies
presently experiencing a burst of star formation can fall in this
region of the diagram and that this burst lasts approximately 10$^8$
years and produce about 10\,\% of the stars of the galaxies.  These
numbers were both derived by the simulation but they perfectly agree
with the measured median mass of the ISOCAM galaxies of
$\sim$5$\times$10$^{10}$ M$_{\odot}$ (from Dickinson et
al. 2003). Indeed in 10$^8$ years and with their median SFR$\sim$ 50
M$_{\odot}$yr$^{-1}$, they produce $\sim$5$\times$10$^{9}$ M$_{\odot}$
of stars, i.e. 10\,\% of the galaxy mass. The mass of newly formed
stars is also consistent with the typical mass of molecular gas found
in local LIRGs. The model used in the Fig.~\ref{FIG:CSFH} predicts
that nearly 50\,\% of present-day stars were formed in a LIRP below
z$\sim$1, hence if this phase only makes up 10\,\% of a galaxy's stars
then the majority if not all of today's galaxies experienced up to
five or even more LIRPs. This suggests that the dense surrounding of
galaxies can trigger successive luminous IR phases, LIRPs, in
galaxies. Finally, the optical morphology of $z\sim$ 0.7 LIRGs derived
from HST-ACS observations (Elbaz et al., in prep, see also Hammer et
al. in these proceedings) shows that less than half of them look like
the major mergers that we see locally in LIRGs. Major mergers might
not be numerous enough to explain such a behavior and passing-by
galaxies might play an important role by triggering strong star
formation events through tidal effects. Hence the local environment of
galaxies, or LEG, might be considered as a better candidate to
understand the origin of distant LIRGs. The cosmic star formation
history is therefore probably strongly dependant on the local density
of galaxies which will also possibly determine their final morphology,
i.e. spirals versus ellipticals, instead of the standard picture of
the merger of two massive disks.

The recently launched Spitzer satellite will provide ideal
observations to check the robustness of this scenario by observing
larger patches of the sky (in particular the SWIRE legacy program),
lowering the effect of cosmic variance, and to extend the study of
luminous IR phases to higher redshifts and lower luminosities (with
the MIPS GTO and GOODS legacy program).

%

\end{document}